\documentclass[aps,prl,twocolumn,floatfix,showpacs,superscriptaddress]{revtex4}

\usepackage{graphicx}
\usepackage{amsmath}
\usepackage{amsfonts}
\usepackage{amssymb}
\usepackage{units}
\usepackage{hyperref}
\usepackage{color}

\newcommand{\degC}{^{\circ}\mathrm{C}}

\begin{document}

\title{Local measurement of the Eliashberg function of Pb islands: enhancement of electron-phonon coupling by quantum well states}

\author{Michael Schackert}
\affiliation{Physikalisches Institut, Karlsruhe Institute of Technology,  Wolfgang-Gaede-Stra{\ss}e 1, 76131 Karlsruhe, Germany}

\author{Tobias M\"arkl}
\affiliation{Physikalisches Institut, Karlsruhe Institute of Technology,  Wolfgang-Gaede-Stra{\ss}e 1, 76131 Karlsruhe, Germany}

\author{Jasmin Jandke}
\affiliation{Physikalisches Institut, Karlsruhe Institute of Technology,  Wolfgang-Gaede-Stra{\ss}e 1, 76131 Karlsruhe, Germany}
 
\author{Martin H\"olzer}
\affiliation{Max-Planck Institut f\"ur Mikrostrukturphysik, Weinberg 2, 06120 Halle, Germany}

\author{Sergey Ostanin}
\affiliation{Max-Planck Institut f\"ur Mikrostrukturphysik, Weinberg 2, 06120 Halle, Germany}

\author{Eberhard K.\,U.\,Gross}
\affiliation{Max-Planck Institut f\"ur Mikrostrukturphysik, Weinberg 2, 06120 Halle, Germany}

\author{Arthur Ernst}
\affiliation{Max-Planck Institut f\"ur Mikrostrukturphysik, Weinberg 2, 06120 Halle, Germany}
\affiliation{Wilhelm-Ostwald-Institut f\"ur Physikalische und Theoretische Chemie,  Linn\'estra{\ss}e 2, Universit\"at Leipzig, 04103 Leipzig, Germany}
  
\author{Wulf Wulfhekel}
\affiliation{Physikalisches Institut, Karlsruhe Institute of Technology,  Wolfgang-Gaede-Stra{\ss}e 1, 76131 Karlsruhe, Germany}

\begin{abstract}
  Inelastic tunneling spectroscopy of Pb islands
  on Cu(111) obtained by scanning tunneling microscopy below \unit[1]{K} provides a
  direct access to the local Eliashberg function of the islands with high energy resolution. 
  The Eliashberg function describes the electron-phonon interaction causing
  conventional superconductivity. The measured Eliashberg function strongly depends on the local thickness of the Pb nanostructures and shows a sharp maximum when quantum well states of the Pb islands come close to the Fermi energy. 
  \textit{Ab  initio} calculations reveal that this is related to enhanced electron-phonon coupling at these thicknesses.
\end{abstract}

\date{\today}
\maketitle

In conventional superconductors (SC) Cooper pairs are formed due to
electron-electron interaction via virtual phonon exchange. While this
concept already lies at the heart of the BCS theory \cite{bardeen_theory_1957} it turned out soon
that the simplifications by the assumptions of the BCS theory are too
crude. Especially when dealing with so-called strong coupling SCs an
extension of the theory is required which was presented by Eliashberg
\cite{eliashberg_interactions_1960}. He took into account that the
electron-phonon interaction is local in space and retarded in time. A
central quantity in this theory is the effective electron-phonon
spectrum $\alpha^2F(\omega)$, which is also called Eliashberg
function. Here $F(\omega)$ is the phonon density of states (DOS) and
$\alpha$ (which actually is $\alpha(\omega)$), is the energy dependent electron-phonon
coupling strength. The Eliashberg theory allows to calculate the
properties for all conventional SCs once $\alpha^2F(\omega)$ is known.

While the experimental determination of $F(\omega)$ can be achieved by
inelastic neutron scattering, this technique provides no immediate access to the full
Eliashberg function since it contains the electron-phonon coupling strength, which is not directly accessible with neutrons.
On the other hand it was shown that the Eliashberg function can be calculated from 
the quasiparticle DOS of a SC extracted from tunneling experiments by inverting the
Eliashberg gap equations \cite{mcmillan_lead_1965}. Although this
method could confirm the validity of the Eliashberg theory it is a
rather indirect way of obtaining $\alpha^2F(\omega)$. Especially in the case of multiple superconducting gaps, the inversion is not unique.

Inelastic tunneling spectroscopy (ITS), however, provides a
direct access to the excitations in a solid. Using scanning
tunneling microscopy (STM), in addition, offers the possibility of
performing ITS with high spatial resolution
\cite{stipe_single-molecule_1998}. Most STM-ITS experiments were
dedicated to molecular vibrations but there are, though rarely,
studies of collective vibrations as well
\cite{vitali_phonon_2004}.
The principle of ITS is based on the opening of an inelastic tunneling channel in parallel to the elastic one
 as soon as the energy of the tunneling electrons
overcomes the threshold for performing an excitation in the
sample leading to an increase of the differential conductivity $\mathrm
dI/\mathrm dU$. 
Since the latter is only of a few
percent it is common to use a lock-in amplifier and record the second
derivative of the tunneling current with respect to the bias voltage,
$\mathrm d^2I/\mathrm dU^2$, which reveals a peak at the positive
voltage corresponding to the energy of the excitation. Due
to the symmetry of electron and hole tunneling, a minimum (dip) is
found at the same voltage on the negative bias side. A dip-peak pair is thus a characteristic fingerprint of an inelastic
tunneling process.  So far, we only considered excitations at discrete
energies. In systems possessing a continuous spectrum of excitations, as in the case of phonons, 
the second derivative of the tunneling current is proportional to
their DOS $F(\omega)$:
\begin{equation}
\left. \frac{\mathrm d^2I}{\mathrm dU^2} \right|_{U=\hbar\omega/e} \propto \rho_t(E_F)\rho_s(E_F)F(\omega)\left|M_{inel}(eU)\right|^2.
\label{eq:IETS}
\end{equation} 
Here $\rho_t(E_F)$ and $\rho_s(E_F)$ are the electronic DOS of the tip and the sample, respectively, and are taken to be
constant around $E_F$. $M_{inel}(eU)$ is the matrix
element for inelastic tunneling which, in the case of phonons, is proportional to the
electron-phonon coupling strength $\alpha$ due to the optical theorem \cite{cohen_quantum_1977}.
In a SC, however, the formation of the gap below the superconducting transition temperature ($T_c$)
leads to a strong energy dependence of $\rho_s(E)$. Since the conductance of a
tunnel junction is directly proportional to the electronic DOS, the gap feature dominates the signal in $\mathrm
d^2I/\mathrm dU^2$. For this reason, superconductivity has to be suppressed. In
principle, there are two possibilities to force a SC into its
normal state even below $T_c$. One is to apply a
magnetic field, the other one is to use the proximity effect of a SC in contact with a normal metal.

In this Letter, we report on the first direct and spatially resolved study of
the Eliashberg function by STM. We used Pb islands on Cu(111) as a model system for
this novel approach and find strong dependencies of the Eliashberg function on the local electronic properties. 

Pb belongs to the so-called strong coupling SCs. The
value of $2\Delta_0/k_B T_c=4.4$ lies far above the BCS value of 3.52
and it is the material with the strongest electron-phonon coupling
among the elemental SCs \cite{buckel_supraleitung_2004} and has
therefore been widely studied in the past.
Pb was among the systems which were investigated in the pioneering
work of Giaever et $\textit{al.}$ who were able to obtain the quasiparticle DOS from planar tunnel
junctions \cite{giaever_tunneling_1962}. Indeed, it was this
measurement from which the Eliashberg function of a SC was obtained
for the first time using the McMillan gap inversion method
\cite{mcmillan_lead_1965}. The Eliashberg function of Pb exhibits two
prominent maxima at around \unit[4]{meV} and \unit[8]{meV}. By
regarding the phonon dispersion relation these features can be
ascribed to van Hove singularities of the transverse and longitudinal phonon modes, respectively
\cite{heid_effect_2010, brockhouse_crystal_1962,
  stedman_phonon-frequency_1967}.

In recent years, especially thin Pb islands attracted much
attention since the vertical confinement of the electrons results in
discrete quantum well states (QWS), which were found to have a
substantial influence on the growth
mode \cite{otero_can_2000,su_correlation_2001}.
When Pb is deposited onto Cu(111) at room
temperature (RT), the growth mode is of Stranski-Krastanov type
\cite{camarero_atomistic_1998}, i.\,e.\ islands start to grow once a wetting layer is
complete. These islands are (111)-oriented nanocrystallites with a
height distribution that is not statistical.  Instead, as was first
observed by Otero \textit{et~al.}, certain "magic" heights are
strongly preferred while some other numbers of monolayers (ML) seem to be
"forbidden" \cite{otero_observation_2002}. The authors show that the
appearance of these preferred heights is directly correlated to the
QWS. A thickness for which a QWS is near the Fermi
level is energetically unfavored and thus, when growing, the system
avoids those numbers of ML.
Moreover, an influence of the QWS on the electron-phonon coupling and on $T_c$ has been
demonstrated
for Pb films on semiconducting Si(111)-($7\times7$)
\cite{zhang_band_2005,guo_superconductivity_2004,eom_persistent_2006}.
Different results were published, in which $T_c$ rises \cite{zhang_band_2005,guo_superconductivity_2004}  or slightly falls  \cite{eom_persistent_2006} with the film thickness superimposed with weak oscillations.
In contrast, Pb islands on Cu(111) are expected to be in the normal
due to the proximity effect of the
metallic substrate \cite{hilsch_zum_1962}. Hence, Pb/Cu(111) should provide an ideal system
to study the phonon excitations by STM-ITS in the normal state and to
resolve the details of the influence of the QWS on the Eliashberg function as well.

The preparation and study of the sample were carried out in UHV ($p
\approx \unit[10^{-10}]{mbar}$) in a home-made low temperature STM
setup \cite{zhang_compact_2011}. In particular, this setup comprises a
Joule-Thomson refrigerator operated with a $^3$He-$^4$He mixture which
allows for a base temperature below \unit[700]{mK}.
The Cu(111) single crystal was cleaned by sputtering with Ar$^+$ ions
at \unit[1.5]{kV} and subsequent annealing at \unit[500]{$\degC$}.
After the sample had cooled down to RT, Pb was
evaporated by electron bombardment and a total amount of
\unit[3.8]{ML} was deposited at a rate of \unit[1.9]{ML/min}. An
electrochemically etched W wire, which was treated by several cycles
of Ar$^+$ sputtering and flash-annealing, was used as STM tip.
The temperature of the sample and the tip was $(0.8\pm0.1)$ K during
the measurements. For ITS the tip was stabilized at \unit[20]{mV}
and \unit[50]{nA} and the spectra were obtained
by measuring the second derivative of the tunneling current using
lock-in technique \cite{stipe_localization_1999} with a modulation
voltage of 400 $\mu$V (RMS) at a frequency of 16.2 kHz. Recording a single
spectrum was set to take \unit[3]{min} but in order to increase the
signal-to-noise ratio 20 spectra were taken successively at the same
point and averaged to one spectrum. These spectra are shown without
any subsequent filtering.

\begin{figure}
  \includegraphics{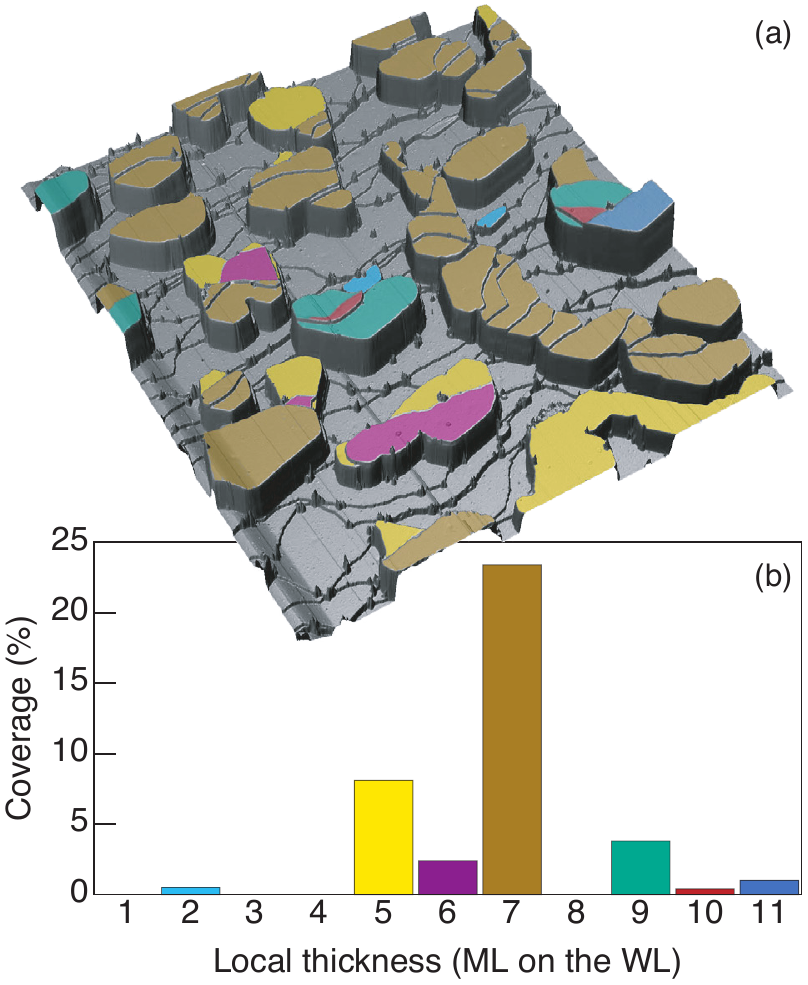}
  \caption
  {(a) STM topography ($\unit[800\times800]{nm^2}$, $\unit[1]{V}, \unit[100]{pA}$) showing Pb islands on the stepped Cu(111) surface. The colored surfaces indicate the local thickness whose distribution is given in (b).}
  \label{fig:growth}
\end{figure}
\begin{figure}
  \includegraphics[width=8.2cm]{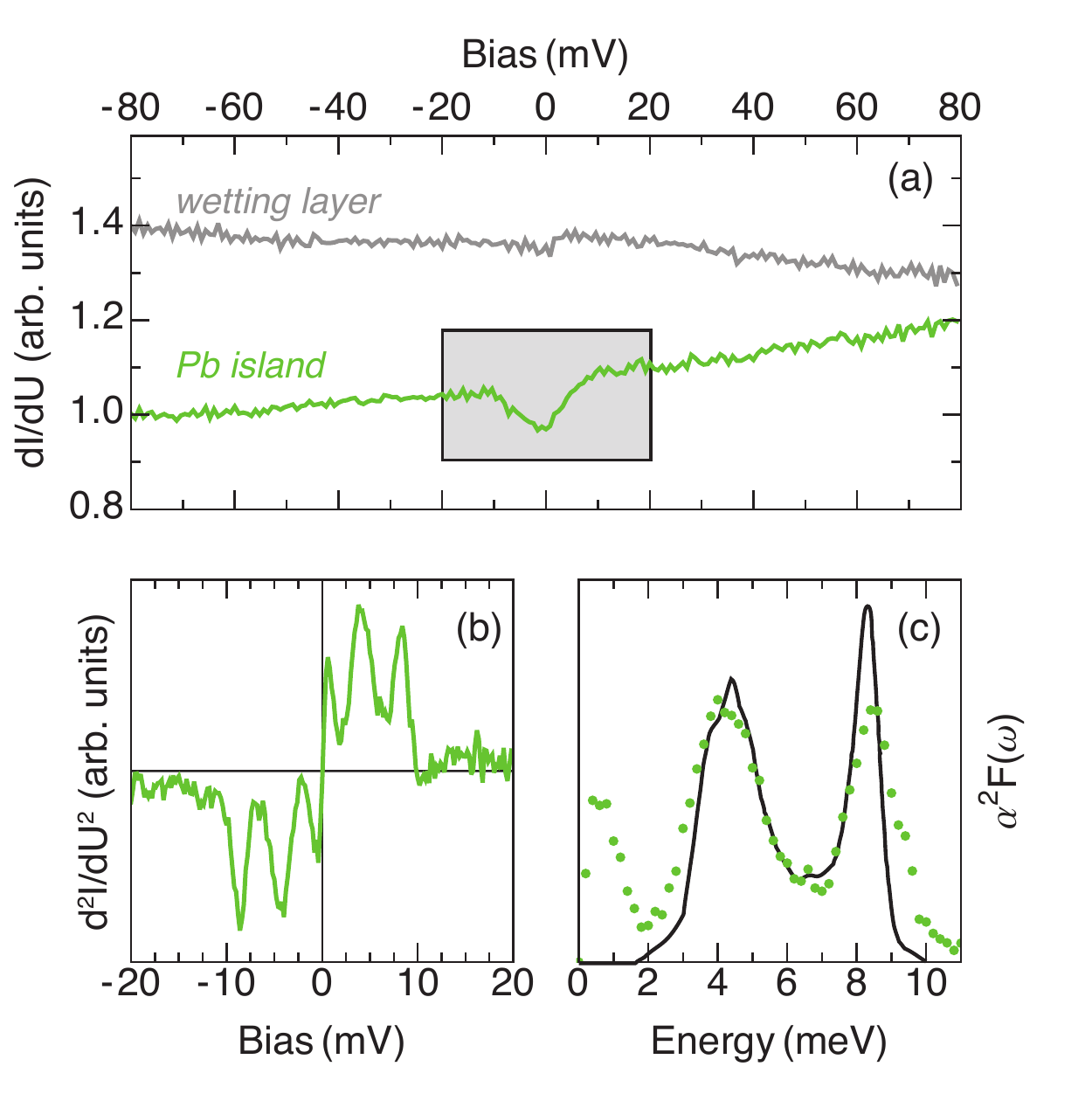}
  \caption{(a) Large energy range tunneling conductance spectra
    (normalized and shifted) of the wetting layer and of a 12 ML thick Pb area.
    While the conductance of the wetting layer is constant, the Pb island
    exhibits is a significant decrease of conductance at zero bias.
    The perfect antisymmetric shape of the $\mathrm d^2I/\mathrm dU^2$
    spectrum (b) taken on the same position in a smaller bias interval
    around $E_F$ doubtlessly reveals the origin of the reduced
    conductance being an inelastic process. (c)Average~ of the positive and negative bias side of
    the ITS spectrum in (b) (dots) in comparison with the previous
    result for $\alpha^2F(\omega)$ (line) \cite{mcmillan_lead_1965}.}
  \label{fig:12MLspec}
\end{figure}
\begin{figure}[b]
  \includegraphics[width=8.2cm]{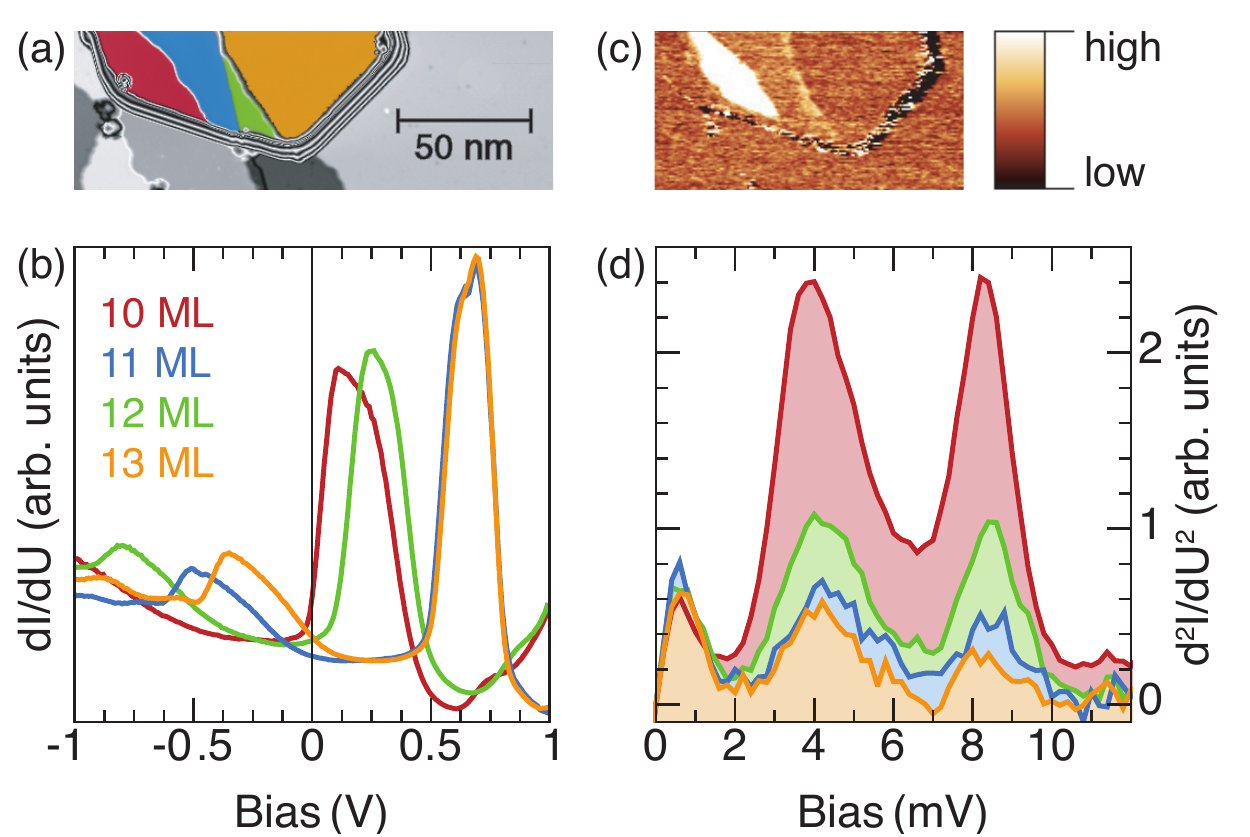}
  \caption{(a) Topographic image of a wedge shaped island. The color
    code of the Pb areas corresponds to the thicknesses given in (b)
    where the QWS obtained from the $\mathrm dI/\mathrm dU$ spectra at
    an arbitrary location on the Pb areas of different heights are
    shown. (d) Average of the positive and negative bias side of the
    ITS spectra recorded at the same locations as in (b). The
    intensity of the $\alpha^2F(\omega)$ increases considerably on the
    10 ML high Pb area as can also be seen from the $\mathrm
    d^2I/\mathrm dU^2$ map of the same island taken at \unit[8]{mV}
    (c).}
  \label{fig:heightdepend}
\end{figure}

Fig.\,\ref{fig:growth}(a) shows a typical STM topography of the
sample. Pb islands of a lateral size of about \unit[100]{nm} grow on
top of the wetting layer on the stepped Cu(111) surface. Most of the islands span
one or more Cu steps and many of them coalesce. In order to determine the
local thickness we considered the height profiles
extracted from the STM topography as well as the QWS \cite{otero_observation_2002}. 
In Fig.\,\ref{fig:growth}(a) the Pb areas are marked with
different colors according to their local thickness whereas
Fig.\,\ref{fig:growth}(b) shows the thickness distribution. Most of the
islands have a constant thickness of 5 or \unit[7]{ML} on top of the wetting layer, which
are thus the "magic heights" for this amount of Pb. As can be seen in
Fig.\,\ref{fig:growth}(a) and \ref{fig:heightdepend}(a), there are,
however, some islands which change their thickness by 1 or \unit[2]{ML}
when growing over a Cu step edge and closely spaced steps may even help
to stabilize unfavored heights, as was already observed before \cite{otero_observation_2002}. In particular, there are only two
rather small regions in Fig.\,\ref{fig:growth}(a) that are 10 ML thick
and \unit[8]{ML} was not observed at all.

We performed $\mathrm dI/\mathrm dU$ spectroscopy on the Pb regions of different
heights as well as on the wetting layer. In Fig.\,\ref{fig:12MLspec} we show
exemplarily the results for a Pb thickness of 12 ML. In contrast to the
constant differential conductance of the wetting layer there is a pronounced
dip around $E_F$ in the case of Pb islands. The same effect was observed for Pb islands on Si(111)-($7\times7$)
\cite{nishio_superconducting_2008,wang_pseudogap_2009}. Since the experiment of Wang \textit{et al.} was performed at
\unit[10]{K}, which is above $T_c$ of Pb, the authors termed this
feature "pseudo gap" as in the field of high temperature superconductivity.
We, however, attribute the dip to inelastic excitations. We present a high resolution $\mathrm d^2I/\mathrm
dU^2$ spectrum of this energy range in Fig.\,\ref{fig:12MLspec}(b). An almost point-symmetric shape is obtained showing two dominant dip-peak pairs at a bias of about \unit[$\pm4.0$]{mV} and
\unit[$\pm8.3$]{mV} as well as one very close to zero bias
(\unit[$\pm0.6$]{mV}). A comparison to the previously determined
Eliashberg function of Pb \cite{mcmillan_lead_1965} is provided in
Fig.\,\ref{fig:12MLspec}(c) and since it reveals an almost perfect agreement there is strong evidence that the ITS data directly yields the Eliashberg function. i.\,e.\ the dip does not indicate a pseudo gap.
Only the feature closest to $E_F$ is not related to
$\alpha^2F(\omega)$ but is due to a zero bias anomaly (ZBA), as already observed in planar tunneling experiments on Pb \cite{wattamaniuk_determination_1971}.

After confirming that STM-ITS is able to measure the Eliashberg
function we focussed on the thickness dependence. In
particular, we investigated a "wedge"-like island
(Fig.\,\ref{fig:heightdepend}(a)) the height of which increases by
\unit[1]{ML} at every underlying step starting from \unit[10]{ML}.
Keeping all parameters constant, we recorded $\mathrm d^2I/\mathrm
dU^2$ spectra (Fig.\,\ref{fig:heightdepend}(d)) on every local thickness and
we were able to find the ZBA and the Eliashberg function in all cases.
While the former stays the same there is a strong thickness dependence
of the intensity of $\alpha^2F(\omega)$. For areas comprising an odd
number of ML it is relatively weak whereas the signal gets stronger
for \unit[12]{ML} and is by far strongest for \unit[10]{ML} as can
also be seen in the $\mathrm d^2I/\mathrm dU^2$ map, i.\,e.\ a map of the local electron-phonon coupling at 8 meV, in
Fig.\,\ref{fig:heightdepend}(c) recorded with the bias set to the
energy of the longitudinal peak. No differences were found for
different locations within an area of a given thickness. 
Note that in all experiments, the current was set to the same value during tip stabilization, such that the differences in the inelastic spectra are not related to differences in the electronic DOS.
Since we
expected that the thickness dependence should be influenced by the
electronic QWS, we measured the first derivative of the tunneling
current in a larger voltage range (Fig.\,\ref{fig:heightdepend}(b)).
In the case of the odd numbers of ML, the highest occupied as well as
the lowest unoccupied states lie roughly \unit[0.5]{eV} away from the
Fermi level. In contrast, the lowest unoccupied QWS of the
\unit[10]{ML} and \unit[12]{ML} thick layers are very close to $E_F$.
The relevant QWS of the former, which exhibited also the highest
$\alpha^2F(\omega)$ intensity, is closest to $E_F$. Thus, our
experiment clearly shows, that electron-phonon coupling can be strongly enhanced (factor 5), if unoccupied QWS are near the Fermi energy.

\begin{figure}
  \includegraphics[width=8.2cm]{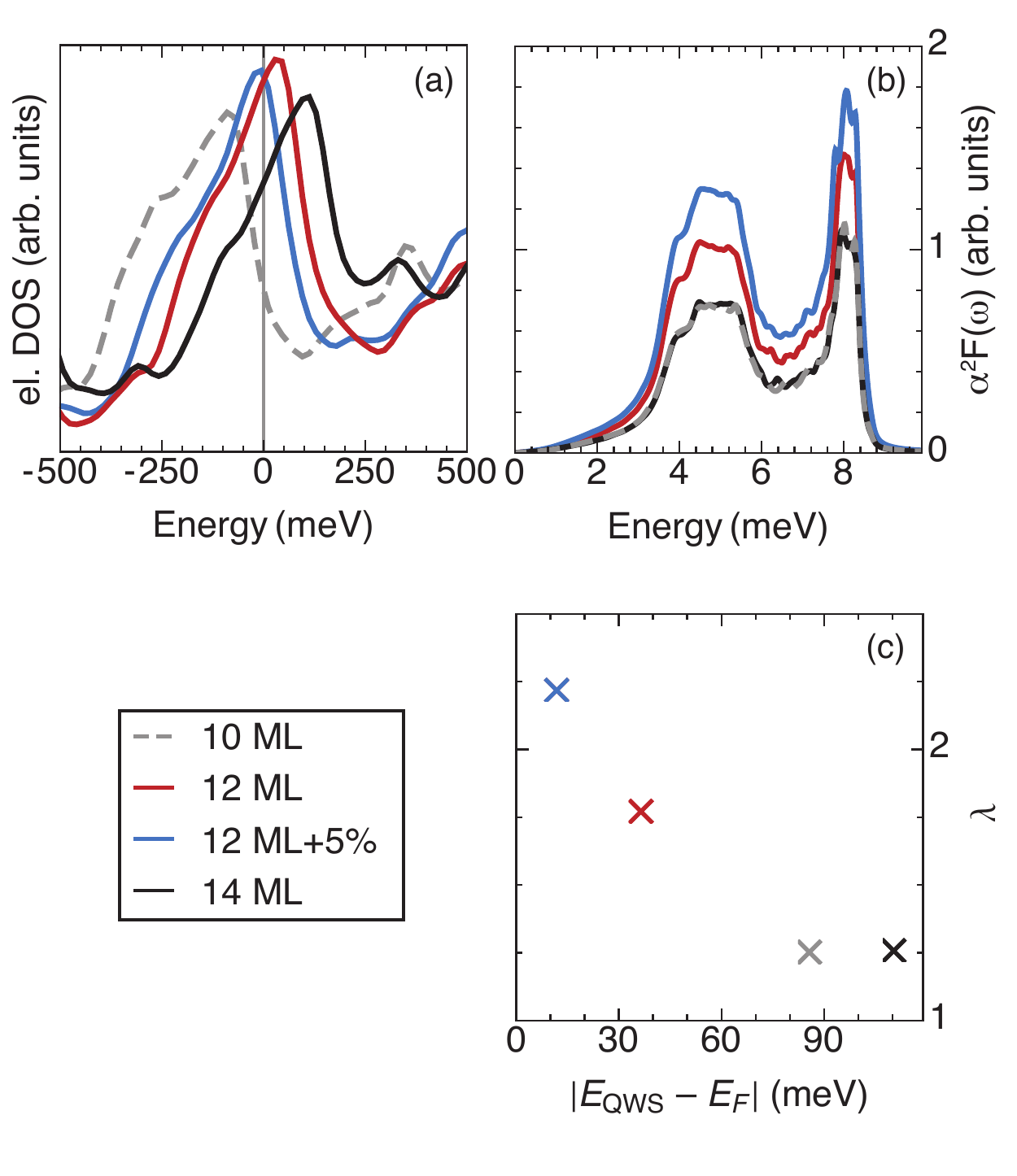}
  \caption{Theoretical calculation. (a) QWS in the electronic DOS (at
    the first vacuum layer). (b)Eliashberg function. (c)~Resulting
    electron-phonon coupling constant $\lambda$.}
  \label{fig:theo}
\end{figure}
To elucidate the experimental finding, we calculated the electronic
structure and phonons for different free standing Pb films.
The structural relaxations and
the phonons were computed using the VASP code, well known for precise
total energy and forces calculations \cite{kresse_efficient_1996}, while the
electronic structure of the thin films was obtained with a
first-principles Green's function method, specially designed for
semi-infinite systems such as surfaces and interfaces
\cite{luders_ab_2001}. The self-consistently calculated Green's functions
and phonons were used to compute the Eliashberg function of the given
systems.

First of all, we found QWS in this systems in qualitative agreement with the experiment. 
The calculated electronic DOS exhibits a QWS closest to
the Fermi level for a thickness of \unit[12]{ML} while for
\unit[10]{ML} and \unit[14]{ML} the QWS lie roughly \unit[0.1]{eV}
below and above $E_F$, respectively (Fig.\,\ref{fig:theo}(a)). The
theoretical results for $\alpha^2F(\omega)$ in Fig.\,\ref{fig:theo}(b)
reveal the same effect as the experiment. While the intensity of
$\alpha^2 F(\omega)$ is the same for 10 and \unit[14]{ML}, it
increases by more than \unit[30]{\%} in the case of \unit[12]{ML}.
Also shown in Fig.\,\ref{fig:theo} is the effect of compressing the
surface of the \unit[12]{ML} slab by \unit[5]{\%}. This compression
shifts the QWS even closer to $E_F$ which again is accompanied by a
further increase of the intensity of $\alpha^2F(\omega)$.
Fig.\,\ref{fig:theo}(c) comprises the resulting electron-phonon coupling constant
$\lambda$, which changes from a value of about 1.25 to 2.2, i.\,e.\ significantly above the bulk value, upon the shift
of the QWS towards $E_F$. Thus, in qualitative agreement to the experiment, the Eliashberg function can be increased by a QWS near the Fermi level. Quantitatively, the experimental enhancement of the electron-phonon coupling is even stronger than theoretically predicted by a factor of 2-3, possibly due to the details of the quantum well states of Pb on Cu(111). 

The mechanism leading to the enhancement can be understood from total energy considerations. The formation of QWS in
thin Pb films leads to an energetically unfavourable condition when a QWS
lies very close to $E_F$. A change of the lattice constant, e.\,g.\ by a phonon, shifts 
the QWS up and down through the Fermi level and leading to repopulation of the electrons with large changes of the electronic energy. Thus, deformations couple more strongly to the
electronic degrees of freedom when a QWS is near the Fermi energy increasing the electron-phonon coupling. 
Supposed that the behavior is similar on Si(111), our findings could
also explain the $T_c$ oscillation with the thickness as it was observed in that system \cite{eom_persistent_2006}.

In conclusion, we demonstrated that, using low temperature
STM-IST, it is possible to directly measure phonon excitations in Pb
islands. The obtained $\mathrm d^2I/\mathrm dU^2$ spectra can be 
unambiguously identified as $\alpha^2F(\omega)$ which allows for the
experimental determination of the latter with the high spatial
resolution of the STM. We found a pronounced dependence of the Eliashberg function
on the thickness of the Pb slabs which, with the aid of
\textit{ab initio} calculations, can be explained by their energetic
stability due to the position of the electronic QWS. This mechanism, if generally applicable, can be used to increase $T_c$ of conventional superconductors by using layered structures with QWS at appropriate energies.

M.S. acknowledges funding by the Karlsruhe House of Young Scientists (KHYS).

\bibliographystyle{apsrev}
\bibliography{./citations} 
\end{document}